# Impact of Two Realistic Mobility Model for Vehicular Safety Applications

Md. Habibur Rahman
Dept. of Computer Science
American International University-Bangladesh
Dhaka, Bangladesh
E-mail: mmmhabib@gmail.com

Mohammad Nasiruddin
Dept. of Computer Science
Univ. Grenoble Alpes
Grenoble, France
E-mail: mohammadnasiruddin@gmail.com

*Abstract*—**Vehicular safety applications intended for VANETs. It can be separated by inter-vehicle communication. It is needed for a vehicle can travel safety with high velocity and must interconnect quickly dependably. In this work, examined the impact of the IDM-IM and IDM-LC mobility model on AODV, AOMDV, DSDV and OLSR routing protocol using Nakagami propagation model and IEEE 802.11p MAC protocol in a particular urban scenario of Dhaka city. The periodic broadcast (PBC) agent is employed to transmit messages between vehicles in case of emergency or collision avoidance for vehicular safety communication. The simulation results recommend numerous concerns such as lower packet drop rate, delay, jitter, route cost and mean-hop is necessary to be measured before developing a robust safety application of VANET.**

*Keywords-VANET; AODV; AOMDV, DSDV; OLSR; IDM-IM; IDM-LC; PBC; IEEE 802.11p; Nakagami*

## I. INTRODUCTION

Vehicular Ad-Hoc Network (VANET) sports vehicles to communicate with each other via vehicle-to-infrastructure (V2I) communications mode and vehicles can interconnect with each other via vehicle-to-vehicle (V2V) communications mode [1]. It offers the timely information to a driver that enables them to predict vehicle collision avoidance, improve congestion, lane change and construction site warning [2]. Vehicular communications systems have drawn fast research attention due to the potential to improve efficiency and safety of car traffic as well as soul and property protection. The IEEE 802.11p standard is used for independent vehicles approaching an intersection use Dedicated Short Range Communications (DSRC) and Wireless Access in a Vehicular Environment (WAVE) to periodically send information such as location, heading and intersection crossing purposes to other vehicles [3]. The signal propagation within DSRC channel can be affected by fading in an urban intersection. To surmount this trouble, various research studies have demonstrated that the nakagami radio propagation model is rather suited for urban scenario [2]-[4]. In VANETs, the routing protocol operation is regarded by various elements such as communication mode, vehicle/node density fluctuations, and guest/vehicle mobility pattern. For realistic vehicular mobility pattern, VanetMobiSim can provide an actual mobility scenario of a specific area [5]. In this work, the periodic broadcast (PBC) agent, is utilized for sending safety message between vehicles in case of emergency or collision avoidance. The IEEE 802.11p MAC protocol and Nakagami radio propagation model is used to get better performance on VANETs. The VanetMobiSim is used to generate realistic vehicular traffic of a particular area of Dhaka city. For experimental evaluations, compared the impact of IDM-IM and IDM-LC on AODV, AOMDV, DSDV and OLSR with respect to several QoS metrics such as Delay, Jitter, Throughput, and other performance metrics such as Average throughput, Normalized Routing Load (NRL), Route Cost, Mean hop and Packet Delivery Ratio (PDR).

## II. RELATED RESEARCH

In [2], the authors investigated that Nakagami propagation model outperforms than TwoRayGround model in an urban scenario for AODV and OLSR routing protocol using IEEE 802.11p MAC protocol. In [4], the authors evaluated the performance of AODV and OLSR routing protocol using Nakagami radio propagation model with IEEE 802.11p MAC protocol. In [5], the author evaluated the performance of VANETs by integrating clustering of different areas and traffic lights via IDM-IM uses AODV and AOMDV protocols with two different types of CBR traffic source connection with existing IEEE 802.11b MAC protocol and TwoRayGround propagation model. In [1], the authors demonstrated the impact of mobility model of IEEE 802.11p performance of vehicular network by investigating certain mobility factors such as relative speed. It holds an important impact on channel access at the MAC layer, brushing off the number of communicating nodes. In [11], the authors have offered that 802.11p gives effective service differentiation mechanism that can be appropriate for the mission-critical ITS application. They evaluated the MAC layer performance without putting on any realistic vehicular mobility model. The operation of IEEE 802.11p MAC sub-layer, principally for the V2V model [1], [11], [12]. Various researchers have used TwoRayGround model instead of Nakagami propagation model for comparing the VANETs performance using different routing protocol, traffic design and various mobility models [5]-[12]. The main drawback of their study is ignoring the safety message transmission in case of emergency or collision avoidance for vehicular communications.

### A. Review of Routing Protocols

Ad-Hoc On-Demand Distance Vector (AODV) routing protocol enables dynamic, on-demand, self-starting, multi-hop routing between participating mobile nodes wishing to build and sustain an ad-hoc network [2], [13], [14]. Ad-Hoc On-Demand Multipath Distance Vector (AOMDV) calculates multiple loop-free and link-disjoint paths, but clients are unaware of the relative movement and positioning [15], [16]. Destination Sequenced Distance Vector (DSDV) is a proactive routing protocol, where each node maintains routing information for each possible destination and can figure out the looping problem and to cope dynamically with network modifications [17]. To optimize the performance of Optimized Link State Routing (OLSR) protocol, Multipoint Relay (MPR) nodes are used for the number of packets broadcasted on the network is minimized [18].

*B. Propagation Model*

The Nakagami radio propagation model is a mathematical general modeling of a radio channel with fading and can provide more configurable parameters to permit a more faithful representation of the wireless communication channel [2]. It can efficiently model the characteristics of different real world scenarios than TwoRayGround model and capable of various scenarios from free space to moderate obstacles to high obstacles can be simulated [4].

*C. Mobility Model*

The VanetMobiSim includes vehicle to infrastructure (V2I) and vehicle to vehicle (V2V) association [5]. It is combined the stop signals, traffic lights and activity based macro-mobility with the funding of human mobility dynamics. IDM-IM support smart intersection management as well as slow down and stop at intersections, or act according to traffic lights, if present [19]. In both situations, it only behaves on the first vehicle on each road, as IDM automatically adjusts the behavior of cars behind the leading car. IDM-LC mobility model provides opportunities for vehicles to alteration lane and overtake among vehicles in the presence of multi-lane roads. These two matters are high by the starter of different lanes such as the parting of traffic flows on dissimilar lanes of the similar road and the overtaking model itself [19].

III. SYSTEM MODEL

Fig. 1 demonstrated a realistic vehicular mobility model in a particular area of Dhaka city using IDM-IM and IDM-LC mobility model to understand the traffic condition of this area. In IDM-IM and IDM-LC, the quantity of interaction with a traffic light is 500. The Traffic light length is 10 seconds. The number of lanes is 2. The maximum number of multi-lane roads are 10. Vehicle length is 5 m. The maximal acceleration of vehicle movement is 0.6 m/s$^2$. The "comfortable" deceleration of vehicle movement is 0.9 m/s$^2$. The minimal distance to a standing node (jam distance) is 1 m. The node's safe time headway is 0.5 s. The step for recalculating movement parameters is 1 s. The maximum stay duration at destination is 6 s. The minimum stay duration at destination is 2 s. The visibility distance of 200 m. The intersections located at the borders of the map is ignored. In IDM-LC, The politeness factor of drivers when changing lane is 0.5. The threshold acceleration for lane change is 0.5 m/s$^2$.

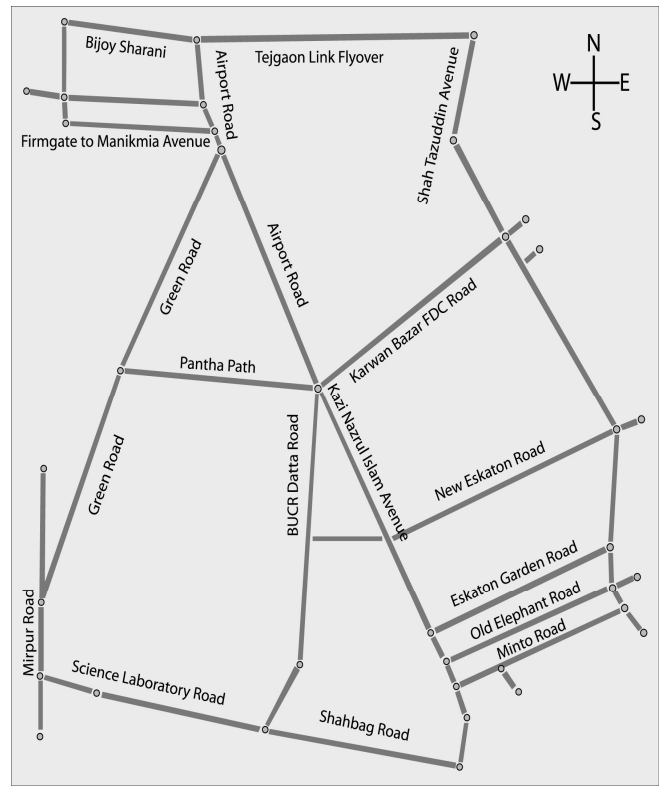

Figure 1. Representation of Vehicular mobility for simulation

*A. Simulation parameter & Quality evaluation*

All nodes use 802.11p MAC operating at 6Mbps. The transmission range is 250 m. For experimental purposes, the simulation area is 1000 X 1000 m$^2$. The real world simulation area is higher than the simulation area used this study. The interface queue length is 50 at each node. The antenna type is Omni-Antenna.

TABLE I. SIMULATION PARAMETERS

| Parameter | Value |
|---|---|
| MAC Type | IEEE 80211p |
| Channel Type | Wireless |
| Mobility Model | As explained in section III. A |
| Simulation Area | 1000 X 1000 m$^2$ |
| Simulation Time | 100 Sec |
| Traffic Model | 40 CBR connection |
| Packet Size | 512 byte |
| No. Of Vehicles | 100 |
| Vehicle Speed | 10 - 80 Km/hr |
| Packet Rate | 4 packets /Sec |
| Radio Propagation Model | Nakagami |
| Routing Protocols | AODV, AOMDV, DSDV, OLSR |

## B. QoS Metrics

*1) Drop:* The packet drop is counted by the total number of packets dropped when a source vehicle transmitted data packet through the network to the destination vehicle. The packet drop ($P_d$) can be counted by Eq. 1.

$$P_d = \sum P_r - \sum P_s \quad (1)$$

Where $P_r$ and $P_s$ are the number of packets received and transmitted respectively.

*2) Throughput:* The throughput is counted as a number of packets that have been efficiently sent to the destination vehicles. The throughput ($T_h$) can be defined as Eq. 2.

$$T_h = \sum N_t \quad (2)$$

Where $N_t$ is the number of data packet bytes in a particular time.

*3) Delay:* A particular packet is transmitted from the source vehicle to the destination and computes the variance between sending times and received times. The delay ($D_i$) can be defined as Eq. 3.

$$D_i = R_t - S_t \quad (3)$$

Where $R_t$ and $S_t$ are the time of packet received and sent.

*4) Jitter:* The jitter is the variance of the packet arrival time. The jitter ($J_i$) can be calculated by Eq. 4.

$$J_i = D_{i+1} - D_i \quad (4)$$

## C. Other Performance Metrics

*1) Average Throughput:* The amount of data sent by the network divided by time period. The average throughput ($A_h$) can be calculated by Eq. 5.

$$A_h = \left(\frac{N_t}{T_n}\right) * \left(\frac{8}{100}\right) \quad (5)$$

Where $T_n$ is the total number of vehicles.

*2) Normalized Routing Load (NRL):* The total number of routing packets transmitted per data packet sent at the destination. The NRL ($N_l$) is calculated by Eq. 6.

$$N_l = \frac{\sum R_p}{\sum P_r} \quad (6)$$

Where $R_p$ is the number of routing packets in layer 2.

*3) Mean Hop:* The total number of control or routing packets forwarded by routing protocol during the simulation to send data packet delivered to the destination. The mean hop ($M_h$) can be defined as Eq. 7.

$$M_h = \frac{\sum P_f}{\sum P_s} \quad (7)$$

Where $P_f$ and $P_s$ are the number routing packet forwarded and sent respectively.

*4) Packet Delivery Ratio (PDR):* The ratio of the data packets sent to the endpoint to those created by the traffic sources. The PDR is calculated by Eq. 8.

$$PDR = \frac{\sum P_r}{\sum P_s} \quad (8)$$

*5) Routing Cost:* It is the ratio of routing bytes to traffic packet bytes. The routing cost ($R_c$) can be calculated by Eq. 9.

$$R_c = \frac{N_r}{N_t} \quad (9)$$

Where $N_r$ and $N_t$ are the number of route & traffic bytes.

## IV. RESULT ANALYSIS

The experiment is implemented using VanetMobiSim and NS 2.35 an Intel (R) Core (TM) i7, Windows 7 and Ubuntu 10.04 platform to measure the impact of two realistic mobility model on four VANET routing protocol in an urban scenario of Dhaka city.

## A. Quantitative Verification

The quantitatively compare the impact of IDM-IM and IDM-LC realistic mobility model using AODV, AOMDV, DSDV, and OLSR routing protocol along with dissimilar parameter of QoS metrics (drop, throughput, delay, jitter) and other performance evaluation metrics (average throughput, normalized routing load, mean-hop, packet delivery ratio, routing cost) for inferring the behavior of CBR packet in the dynamic network simulation scenario. The simulation result is presented in the following TABLE II-TABLE XVII and Fig. 2-Fig. 9.

TABLE II. NUMBER OF PACKET DROP FOR AODV

| Packet type | Mobility model | Total sent packets | Total received packets | Totally dropped packets |
|---|---|---|---|---|
| CBR | IDM-IM | 13153 | 5594 | 7559 |
| | IDM-LC | 13091 | 5506 | 7585 |

TABLE III. NUMBER OF PACKET DROP FOR AOMDV

| Packet type | Mobility model | Total sent packets | Total received packets | Totally dropped packets |
|---|---|---|---|---|
| CBR | IDM-IM | 13163 | 5568 | 7595 |
| | IDM-LC | 13112 | 5609 | 7503 |

TABLE IV. NUMBER OF PACKET DROP FOR DSDV

| Packet type | Mobility model | Total sent packets | Total received packets | Totally dropped packets |
|---|---|---|---|---|
| CBR | IDM-IM | 13047 | 3902 | 9145 |
| | IDM-LC | 9408 | 3065 | 6343 |

TABLE V. NUMBER OF PACKET DROP FOR OLSR

| Packet type | Mobility model | Total sent packets | Total received packets | Totally dropped packets |
|---|---|---|---|---|
| CBR | IDM-IM | 13098 | 6166 | 6932 |
| | IDM-LC | 13068 | 5981 | 7087 |

TABLE VI. THROUGHPUT FOR AODV

| Packet type | Mobility model | Total sent throughput (kbps) | Total received throughput (kbps) |
|---|---|---|---|
| CBR | IDM-IM | 6734336 | 2864128 |
| | IDM-LC | 6702592 | 2819072 |

TABLE VII. THROUGHPUT FOR AOMDV

| Packet type | Mobility model | Total sent throughput (kbps) | Total received throughput (kbps) |
|---|---|---|---|
| CBR | IDM-IM | 6739456 | 2850816 |
| | IDM-LC | 6713344 | 2871808 |

TABLE VIII. THROUGHPUT FOR DSDV

| Packet type | Mobility model | Total sent throughput (kbps) | Total received throughput (kbps) |
|---|---|---|---|
| CBR | IDM-IM | 6680064 | 1997824 |
| | IDM-LC | 4816896 | 1569280 |

TABLE IX. THROUGHPUT FOR OLSR

| Packet type | Mobility model | Total sent throughput (kbps) | Total received throughput (kbps) |
|---|---|---|---|
| CBR | IDM-IM | 6706176 | 3156992 |
| | IDM-LC | 6690816 | 3062272 |

TABLE X. PDR, DROP & AVG. THROUGHPUT FOR AODV

| Packet type | Mobility model | PDR (%) | Drop (%) | Avg. throughput (kbps) |
|---|---|---|---|---|
| CBR | IDM-IM | 42.53 | 57.47 | 229.14 |
| | IDM-LC | 42.06 | 57.94 | 225.53 |

TABLE XI. PDR, DROP & AVG. THROUGHPUT FOR AOMDV

| Packet type | Mobility model | PDR (%) | Drop (%) | Avg. throughput (kbps) |
|---|---|---|---|---|
| CBR | IDM-IM | 42.30 | 57.70 | 228.09 |
| | IDM-LC | 42.78 | 57.22 | 229.75 |

TABLE XII. PDR, DROP & AVG. THROUGHPUT FOR DSDV

| Packet type | Mobility model | PDR (%) | Drop (%) | Avg. throughput (kbps) |
|---|---|---|---|---|
| CBR | IDM-IM | 29.91 | 70.09 | 159.83 |
| | IDM-LC | 32.58 | 67.42 | 147.72 |

TABLE XIII. PDR, DROP & AVG. THROUGHPUT FOR OLSR

| Packet type | Mobility model | PDR (%) | Drop (%) | Avg. throughput (kbps) |
|---|---|---|---|---|
| CBR | IDM-IM | 47.08 | 52.92 | 252.57 |
| | IDM-LC | 45.77 | 54.23 | 245.00 |

TABLE XIV. NRL, ROUTE COST & MEAN HOP FOR AODV

| Packet type | Mobility model | NRL | Route cost | Mean hop |
|---|---|---|---|---|
| CBR | IDM-IM | 49.745 | 0.057 | 1.793 |
| | IDM-LC | 50.508 | 0.056 | 1.843 |

TABLE XV. NRL, ROUTE COST & MEAN HOP FOR AOMDV

| Packet type | Mobility model | NRL | Route cost | Mean hop |
|---|---|---|---|---|
| CBR | IDM-IM | 11.076 | 0.015 | 1.171 |
| | IDM-LC | 10.595 | 0.014 | 1.165 |

TABLE XVI. NRL, ROUTE COST & MEAN HOP FOR DSDV

| Packet type | Mobility model | NRL | Route cost | Mean hop |
|---|---|---|---|---|
| CBR | IDM-IM | 1.048 | 0.003 | 1.000 |
| | IDM-LC | 0.993 | 0.003 | 1.000 |

TABLE XVII. NRL, ROUTE COST & MEAN HOP FOR OLSR

| Packet type | Mobility model | NRL | Route cost | Mean hop |
|---|---|---|---|---|
| CBR | IDM-IM | 2.234 | 0.006 | 1.000 |
| | IDM-LC | 2.273 | 0.006 | 1.000 |

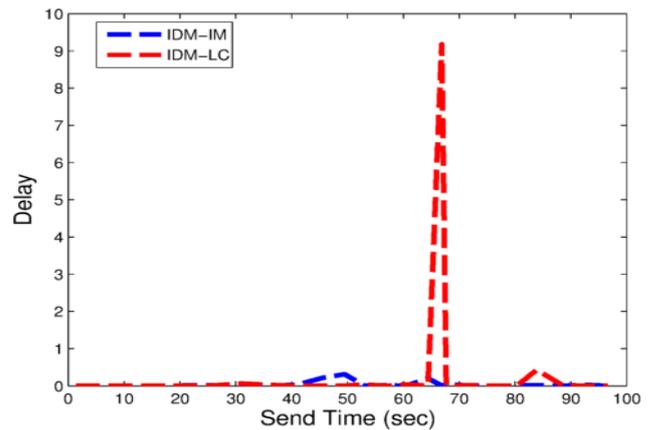

Figure 2. Delay for AODV

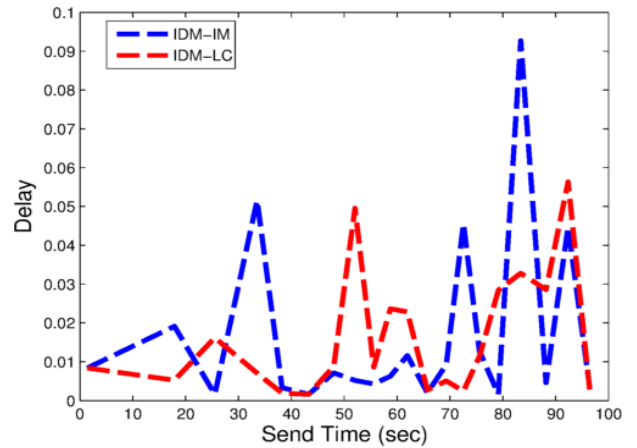

Figure 3. Delay for AOMDV

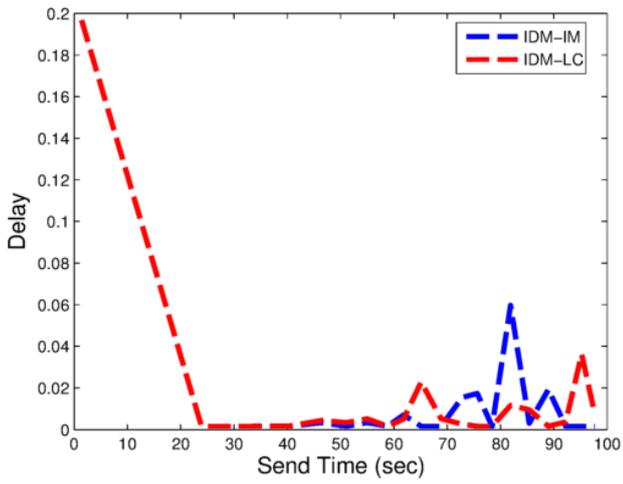

Figure 4. Delay for DSDV

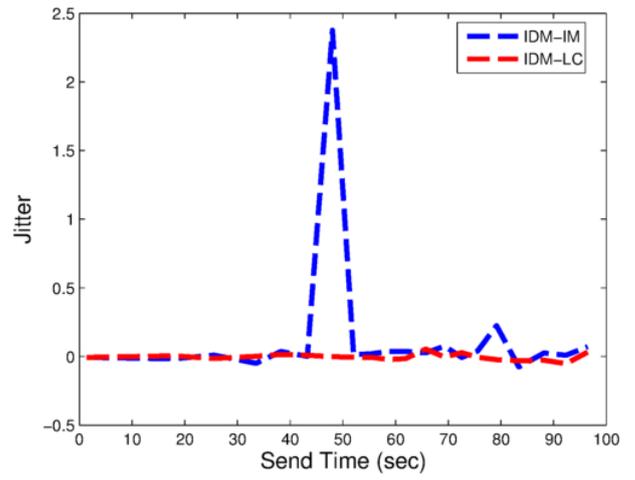

Figure 7. Jitter for AOMDV

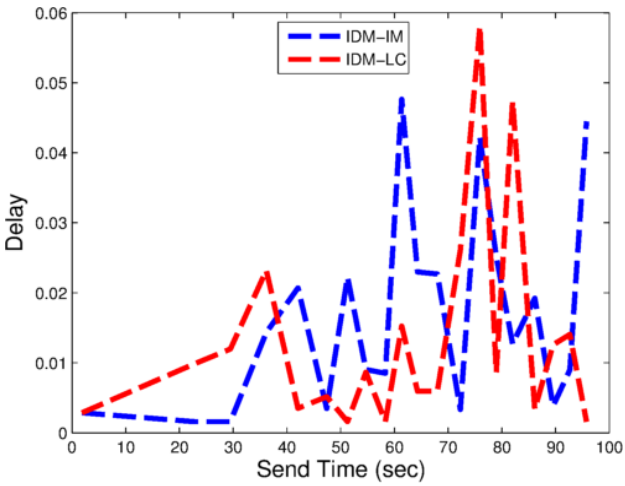

Figure 5. Delay for OLSR

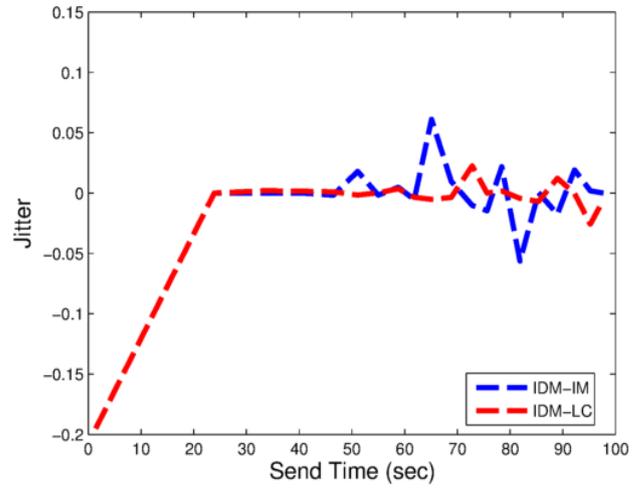

Figure 8. Jitter for DSDV

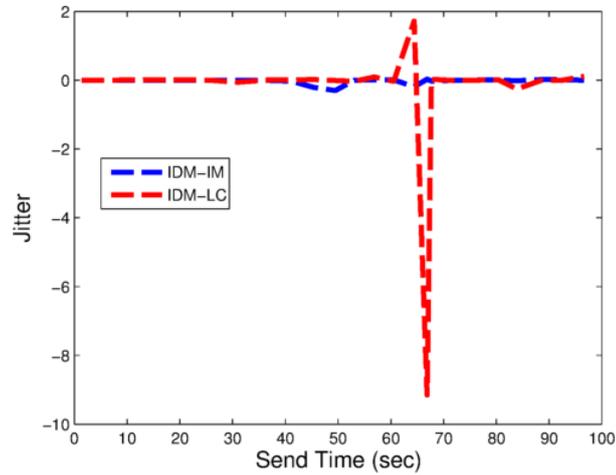

Figure 6. Jitter for AODV

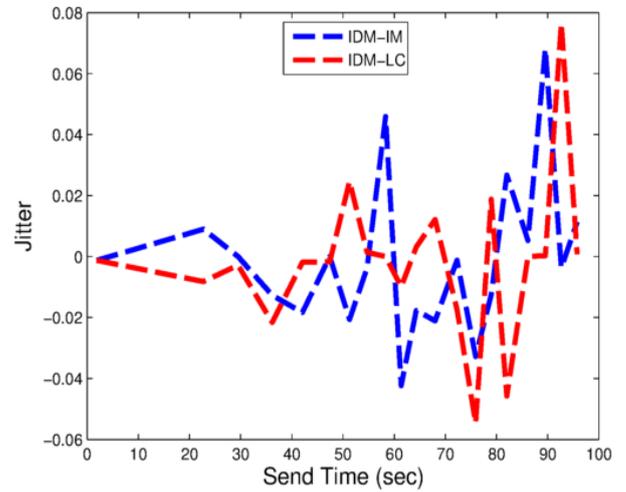

Figure 9. Jitter for OLSR

In TABLE II, TABLE V, TABLE X and TABLE XIV have shown that IDM-IM outperforms than IDM-LC model for the computed mobility impact of AODV in case of Drop rate (57.47%, 57.94%), PDR (42.53%, 42.06%) and Average Throughput (229.14 kbps, 225.53 kbps) respectively. In

TABLE III, TABLE VII, TABLE XI, and TABLE XV has shown that IDM-LC performs better than IDM-IM for AOMDV protocol in case of Drop rate (42.30%, 42.78%), PDR (57.70%, 57.22%) and Average Throughput (228.09 kbps, 229.75 kbps). In TABLE 4, TABLE 8, TABLE XII and TABLE XVI have shown that IDM-LC is performing better than IDM-IM for DSDV in case of Drop rate (70.09%, 67.42%) and the PDR (29.91%, 32.58%). In TABLE V, TABLE IX, TABLE XIII and TABLE XVII have shown that IDM-IM model is performing better than IDM-LC for OLSR in case of Drop rate (52.92%, 54.23%), PDR (47.08%, 45.77%) and Average Throughput (252.57 kbps, 245.00 kbps). In TABLE XIV shows that IDM-IM outperforms among IDM-LC for AODV with respect to NRL and Mean hop. In TABLE XV shows IDM-LC outperforms among IDM-IM for AOMDV routing protocol with respect to NRL and Mean hop. In TABLE XVI and TABLE XVII shows IDM-LC outperforms among IDM-IM for DSDV and OLSR routing protocols in case of NRL. In Fig. 2 and Fig. 6 has shown that the calculation of delay and jitter for AODV, IDM-IM performs better than IDM-LC. In Fig. 3, Fig. 4, Fig. 5, Fig. 7, Fig. 8 and Fig. 9 shows that IDM-IM and IDM-LC can perform well in a certain time not always for AOMDV, DSDV and OLSR routing protocols in case of delay and jitter.

## V. CONCLUSIONS

In this work, two realistic mobility model (IDM-IM and IDM-LC) are used to show their impact on four routing protocols (AODV, AOMDV, DSDV and OLSR) using Nakagami propagation and IEEE 802.11p MAC layer with respect to QoS and other performance metrics in an urban scenario. The safety message is periodically broadcasted using a PBC agent to avoid accident or collision avoidance for each vehicle. In the simulation result, it is distinctly suggested that four routing protocols, and two mobility models were not up to the mark for each of the parameters of performance metrics/QoS metrics towards the development of realistic vehicular safety applications.